\documentclass[fleqn,usenatbib]{mnras}

\usepackage{newtxtext,newtxmath}

\usepackage[T1]{fontenc}

\DeclareRobustCommand{\VAN}[3]{#2}
\let\VANthebibliography\thebibliography
\def\thebibliography{\DeclareRobustCommand{\VAN}[3]{##3}\VANthebibliography}

\usepackage{graphicx}	
\usepackage{amsmath}	
\usepackage[section]{placeins}	
\usepackage{wrapfig}        
\usepackage{bm}
\usepackage{pdflscape}
\usepackage{pgf}
\usepackage{makecell}

\DeclareMathOperator\sign{sgn}
\mathchardef\mhyphen="2D

\title[$\theta\mhyphen\theta$ VLBI Methods]{High Resolution VLBI Astrometry of pulsar scintillation screens with the $\theta\mhyphen\theta$ Transform}

\author[D. Baker et al.]{
Daniel Baker,$^{1,2,3}$\thanks{E-mail: dbaker@asiaa.sinica.edu.tw}
Walter Brisken,$^{4}$
Marten H. van Kerkwijk,$^{5}$
Rik van Lieshout$^{1}$
Ue-Li Pen,$^{1,2,5}$
\\
$^{1}$Academia Sinica Institute of Astronomy and Astrophysics (ASIAA), No. 1, Sec. 4, Roosevelt Rd., Taipei 10617, Taiwan\\
$^{2}$Canadian Institute for Theoretical Astrophysics, University of Toronto, 60 St. George Street, Toronto, ON M5S 3H8, Canada\\
$^{3}$Department of Physics, University of Toronto, 60 St. George Street, Toronto, ON M5S 1A7, Canada\\
$^{4}$National Radio Astronomy Observatory, Socorro, NM 87801, USA\\
$^{5}$ Department of Astronomy and Astrophysics, University of Toronto, 50 St. George Street, Toronto, ON M5S 3H4, Canada\\
}

\date{Accepted XXX. Received YYY; in original form ZZZ}

\pubyear{2023}

\begin{document}
\label{firstpage}
\pagerange{\pageref{firstpage}--\pageref{lastpage}}
\maketitle

\begin{abstract}
The recent development of $\theta\mhyphen\theta$ techniques in pulsar scintillometry has opened the door for new high resolution imaging techniques of the scattering medium. By solving the phase retrieval problem and recovering the wavefield from a pulsar dynamic spectrum, the Doppler shift, time delay, and phase offset of individual images can be determined. However, the results of phase retrieval from a single dish are only known up to a constant phase rotation, which introduces extra parameters when doing astrometry using Very Long Baseline Interferometry. We present an extension to previous $\theta\mhyphen\theta$ methods using the interferometric visibilities between multiple stations to calibrate the wavefields. When applied to existing data for PSR B0834+06 we measure the effective screen distance and lens orientation with five times greater precision than previous works.
\end{abstract}




\section{Introduction}
Scattering of radio emission by the Interstellar Medium (ISM) is a general nuisance to the radio astronomy community. For imaging experiments such as the Event Horizon Telescope (EHT) it leads to distortions of the final image \citep{Zhu2019}. For Pulsar Timing Arrays (PTAs), it increases timing noise and can even masquerade as a gravitational wave signal \citep{Main2020}. In order to the escape these problems, PTAs use high frequency observations where the effects of scattering are reduced. However, pulsar brightness falls off at higher frequencies and so the removal of scattering effects comes at the cost of the number of pulsars available for the array. An understanding of scattering structures, which we will refer to as lenses, in the ISM, particularly those affecting pulsars, would be a boon to the field. Fortunately, pulsar observations provide an excellent source of information about the ISM.
As effectively point sources, pulsars exhibit scintillation behaviour that allows us to probe small scale structure in the ISM. Variations in free electron density result in the formation of multiple images of the pulsar on the sky whose interference pattern can be observed. The standard representation of this phenomenon is the observed intensity as a function of both time and frequency known as the dynamic spectrum.
Originally believed to be the result of turbulence in the ISM, our understanding of pulsar scintillation underwent a fundamental shift with the discovery of parabolic scintillation arcs by \cite{Stinebring2001}. These features are seen in the 2D Fourier transform of the dynamic spectrum, known as the conjugate spectrum, or its magnitude squared, known as the secondary spectrum; and are indicative of the lensed images falling along a linear feature in the sky. In some systems, the main parabolic arc can be seen to consist of a collection of inverted arclets of the same curvature extending downwards from points on the main parabola. Points along the main parabola are interpreted as the interference of scattered images with a "line of sight" image, while the inverted arclets are caused by the interference of pairs of scattered images.

One proposal for the source of these parabolic structures is corrugated reconnection sheets \citep{Pen2014}. These structures are expected to exist along magnetic domain boundaries in the ISM and support surface waves. When the sheet is nearly aligned with the line of sight to the pulsar, grazing light at wave crests will be strongly refracted. If the sheet possesses multiple crests, the resulting images will appear nearly linear on the sky analogously to light reflecting off a wavy lake. ~\cite{Simard2018} expand on this model to predict the evolution of these images over both time and frequency.

In order to test this model, one would like to produce high resolution astrometric images of the screen. For PSR B0834+06, this has been achieved by \cite{Brisken2009} using very-long-baseline interferometry (VLBI) measurements to produce $100 \mu \text{as}$ resolution images of the screen. Since the points in the conjugate spectrum along the main arc, or alternatively the apexes of the inverted arclets, represent the interference of images with the line of sight, their interferrometric phases encode information about their angular offset. The first step in forming an image is to identify these points in the conjugate spectrum. Since the inverted arclets are extended, and may be densely packed, this identification is nontrivial. By solving the phase retrieval problem and restoring the underlying wavefield at each dish, individual images can be identified as isolated points in its Fourier transform: the conjugate wavefield.

The first application of a method for solving the phase retrieval problem in pulsar scintillometry is presented in \cite{Walker2005}. This approach iteratively adds new images to the the conjugate wavefield to build up an approximate solution similar to the CLEAN algorithm. However, the invention of the $\theta\mhyphen\theta$ transform by \cite{Sprenger2020} opens up a new approach to phase retrieval. Under the assumption of a thin one dimensional lens whose features are fixed relative to each other, this transformation maps the conjugate spectra into a space where the inverted arclets form a grid of horizontal and vertical lines. \cite{Baker2021} show how the complex magnifications of each image can be found using the eigenvectors of the transformed spectrum and how these can be mapped back to reproduce the wavefield. Unfortunately, since phases recovered using a single dish can only be found up to some unknown phase rotation as rotating a solution by any constant phase will leave the dynamic spectrum unchanged, there will in general be some unknown additional phase offset between the dishes. This unknown phase can be corrected for, but requires an additional fit parameter when determineing screen properties such as the orientation. 

In this paper we present a method for using the additional information contained in the visibilites from VLBI observations to fix the relative phases  of the recovered wavefields and allow for astrometric imaging. Additionally, this method makes use of the gernally higher signal to noise of visiblites involving one small and one large dish, when compared to the small dish alone, to recover the wavefield at stations that might no be recoverable using only the dynamic spectrum. The application of this method is demonstrated on the observation of B0834+06 of \cite{Brisken2009}. This data is particularly interesting as the presence of two discrete collections of images: a traditional scintillation arc, whose apex is at the origin and has the same curvature as its inverted arclets; and a collection of arclets offset from that main arc at a time delay of approximately 1 ms, suggests lensing by multiple screens. The theory behind this technique is described in Sec.~\ref{sec:theory}. In Sec.~\ref{sec:Data} we describe the observations used to test this method as well as their preparation for analysis. In Sec.~\ref{sec:phase_ret} we discuss the recovered wavefields. In Sec.~\ref{sec:lens_params} we show how the geometry of the lenses, including effective distances, can be determined. Finally, Sec.~\ref{sec:astromet} presents an astrometric image produced from our analysis

\section{VLBI with $\theta\mhyphen\theta$}
\label{sec:theory}
\begin{figure}
    \centering
    \includegraphics{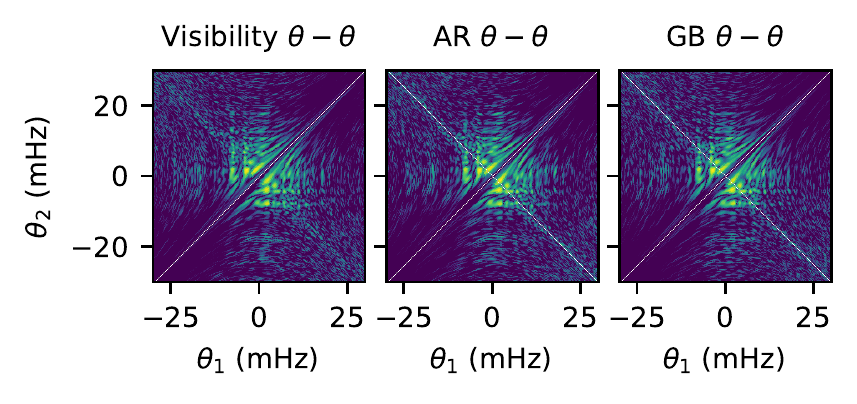}
    \caption{Amplitudes of the $\theta\mhyphen\theta$ spectra from the same portion of the data from dynamic spectra of Arecibo and Green Bank as well as the visibility between the two.}
    \label{fig:vlbi_single_thths}
\end{figure}

\begin{figure}
    \centering
    \includegraphics{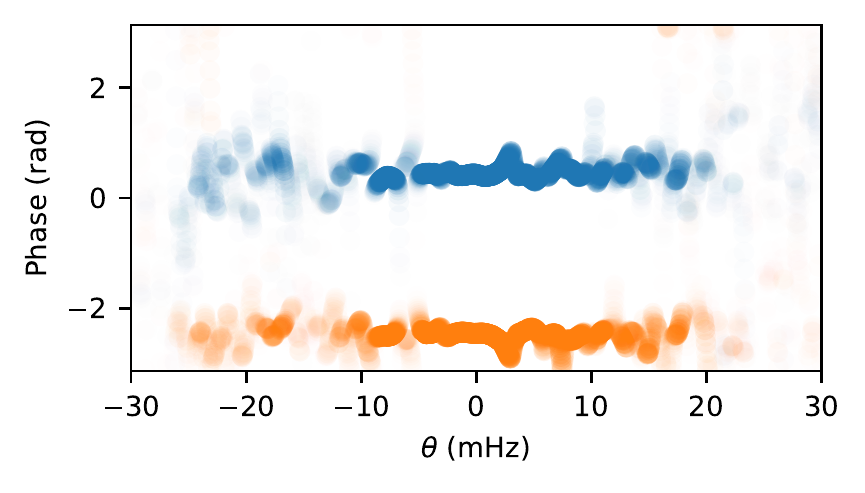}
    \caption{Relative phase between the magnifications measured on a single dish and from the visibility for Arecibo (blue) and Green Bank (orange). The transparency of each point shows the amplitude relative to the peak magnification for each dish. Both dishes show constant (and unrelated) phase offsets between the two methods due to the non uniqueness of the $\theta\mhyphen\theta$ solution under phase rotations.}
    \label{fig:visibility_veccomps}
\end{figure}

For single dish observations, $\theta\mhyphen\theta$ methods have been shown to be able to recover the wavefields underlying pulsar dynamic spectra. However, since this amounts to undoing the convolution of the wavefield with itself there is an unknown overall phase rotation in the result. Since interferometry relies on the relative phases between stations, it is insufficient to independently recover the fields for each station. In order to avoid the problem we must simultaneously perform the recovery at all stations and include information about their relative phases. To this end, we consider the visibility between any two stations. A breakdown of the quantities discussed in this approach is shown in Table~\ref{tab:definitions}, which serves as an extension of the table in \cite{Simard2019}. For dynamic wavefields at the two stations $W_i(\nu,t)$ and $W_j(\nu,t)$, the visibility is given by
\begin{equation}
    V_{i,j}\left(\nu,t\right)=W_i\left(\nu,t\right) W_j^*\left(\nu,t\right)
\end{equation}
and so, denoting Fourier transforms with a tilde,
\begin{equation}
    \widetilde{V}_{i,j}\left(\tau,f_D\right)=\widetilde{W}_i\left(\tau,f_D\right) *\widetilde{W}_j^*\left(-\tau,-f_D\right)
\end{equation}
where $f_D$ and $\tau$ are the conjugate variables to time and frequency respectively, and $\widetilde{W}_i$ and $\widetilde{W}_j$ are the conjugate wavefields. For each dish, the conjugate spectrum is given by the convolution
\begin{equation}
    \widetilde{I}_i\left(\tau,f_D\right)=\widetilde{W}_i\left(\tau,f_D\right) *\widetilde{W}_i^*\left(-\tau,-f_D\right)
\end{equation}
or in $\theta\mhyphen\theta$ space as the outer product
\begin{equation}
    \mathbfss{I}_i = \bm{\mu}_i \otimes \bm{\mu}^*_i
\end{equation}
where $\bm{\mu}_i$ is the complex response vector giving the magnification and phase rotation of points along the lens as described in \cite{Baker2021}. Similarly, the visibility in the $\theta\mhyphen\theta$ space becomes
\begin{equation}
    \mathbfss{V}_{i,j}=\bm{\mu}_i \otimes \bm{\mu}^*_j
\end{equation}
It follows that most of the information for the magnifications from each dish can be recovered from the visibility alone. To demonstrate this, we consider a simultaneous chunk of data from the Arecibo (AR) and Green Bank (GB) dynamic spectra (see \ref{sec:Data} as well as their visibility. Using the fitted curvature for the parabolic arc as described in \cite{Baker2021}, we generate their $\theta\mhyphen\theta$ spectra in Fig.~\ref{fig:vlbi_single_thths}
Unlike the case of single dish $\theta\mhyphen\theta$, the visibility $\theta\mhyphen\theta$ is the outer product of two vectors that differ by more than just complex conjugation. In this case, we cannot recover the vectors from an eigenvector decomposition and instead use the Singular Value Decomposition (SVD) which decomposes a matrix $M$ into singular valuesm $S_m$, and two sets of vectors $\bm{A}_m$ and $\bm{B}_m$ such that
\begin{equation}
    M=\sum_{m=1}^{n} S_m \bm{A}_m \otimes \bm{B}_m
    \label{eq:svd_full}
\end{equation}
As in the case of the single dish eigenvector decomposition, only the first mode contributes to the signal and so
\begin{equation}
\begin{split}
    \mathbfss{I}_i&=S_i \bm{a}_i \otimes \bm{b}_i \\
    \mathbfss{V}_{i,j}&=S_{i,j} \bm{a}_{i,j} \otimes \bm{b}_{i,j}
\end{split}
\label{eq:svd_def}
\end{equation}
where $\bm{a}$ and $\bm{b}$ are vectors and $S$ is the singular value. Combining this with our definition of the $\theta\mhyphen\theta$ spectra for single dishes and visibilities, it follows that
\begin{equation}
    \bm{\mu}_i = \bm{a}_{i} = \bm{a}_{i,j} = \bm{b}_{j,i}^*
\end{equation}
It is worth noting at this stage that multiplying $\bm{a}_{i,j}$  while dividing $\bm{b}_{i,j}$ by the same constant will not change our final model. The relative amplitudes of the $\bm{\mu}_i$ and $\bm{\mu}_j$ cannot be determined from the visibilities alone. As such we restrict ourselves to examining the phases. For our example, the phases of $\bm{a}_{AR}$ and $\bm{a}_{AR,GB}$ should agree up to a constant offset, and similarly for $\bm{a}_{GB}$ and $\bm{b}_{AR,GB}^*$.
Fig~\ref{fig:visibility_veccomps} shows the difference in phase of the two approaches for each dish. As expected, both dishes are offset by --different-- constant phases. This difference arises from the random phase associated with the single dish recoveries since the relative phases between Arecibo and Green Bank are forced by the visibility.
In theory, the visibilities alone could be used to determine the relative phases for each baseline in an array one at a time. However, a more elegant solution exists allowing us to find the relative phases and amplitudes at all dishes simultaneously.
If we define the concatenated response vector for a VLBI observation with n stations as
\begin{equation}
    \bm{\mu} = \bm{\mu}_1 \oplus \bm{\mu}_2 \oplus \cdots \oplus \bm{\mu}_n
    \label{eq:concat_u}
\end{equation}
then the outer product $\bm{\mu} \otimes \bm{\mu}$ will be given by the block matrix
\begin{equation}
\bm{\mu} \otimes \bm{\mu}^* =
    \begin{pmatrix}
    \mathbfss{I}_1 & \mathbfss{V}_{1,2} & \cdots & \mathbfss{V}_{1,n}\\

    \mathbfss{V}_{1,2}^\dagger & \mathbfss{I}_2 & \cdots & \mathbfss{V}_{2,n}\\
    \vdots & \vdots & \ddots & \vdots \\
    \mathbfss{V}^\dagger_{1,n} & \mathbfss{V}^\dagger_{2,n} & \cdots & \mathbfss{I}_n
    \end{pmatrix}
\end{equation}
where $\dagger$ denotes the Hermitian conjugate
We can now recover $\bm{\mu}$, as for the single dish case, by taking the dominant eigenvector. Here we have returned to using the eigenvector decomposition as the vectors in the outer product are a complex conjugate pair. The unknown phase rotation for the individual $\bm{\mu}_i$ is now held fixed, and the relative phases can be determined.
\begin{landscape}
    \begin{table}
            \caption{Definitions of common quantities in scintillometry using VLBI and their relations to the wavefield.}
            \centering
            \begin{tabular}{||c|c|l||}
                \hline
                Parameter & Symbol & Description
                \\
                \hline
                \multicolumn{3}{c}{Single Dish Quantities}
                \\
                \hline
                Dynamic wavefield & $W_{i}\left(\nu,t\right)$ & \makecell[l]{The dynamic frequency response of the interstellar medium. A Fourier transform\\along frequency gives the impulse response in each time bin.}
                \\
                Dynamic spectrum & $I_{i}\left(\nu,t\right) = |W_{i}\left(\nu,t\right)|^2$ & The observed intensity of the pulsar at station i, at frequency $\nu$ and time $t$.
                \\
                Conjugate wavefield & $\widetilde{W}_{i}\left(\tau,f_D\right)$ & The Fourier transform of the dynamic wavefield for station i.
                \\
                Conjugate spectrum & $\widetilde{I}_{i}\left(\tau,f_D\right) = \left(\widetilde{W}_{i}* \widetilde{W}^*_{i}\right)\left(\tau,f_D\right)$  & The Fourier transform of the dynamic spectrum at station i. \\
                Secondary spectrum & $S_{Ii,i}\left(\tau,f_D\right) = |\widetilde{I}_{i}\left(\tau,f_D\right)|^2$ & The two dimensional power spectrum of the dynamic spectrum at station i.
                \\
                \hline
                \multicolumn{3}{c}{VLBI Quantities}
                \\
                \hline
                Cross conjugate wave-spectrum & $S_{Wi,j}\left(\tau,f_D\right) = \widetilde{W}_{i}\left(\tau,f_D\right)\widetilde{W}^{*}_{j}\left(\tau,f_D\right)$ & The cross-correlation of the dynamic wavefields at stations i and j, in the Fourier domain.
                \\
                Intensity cross secondary spectrum 	 & $S_{Ii,j}\left(\tau,f_D\right) = \widetilde{I}_{i}\left(\tau,f_D\right)\widetilde{I}^{*}_{j}\left(\tau,f_D\right)$ & The cross-correlation of the dynamic spectra at stations i and j, in the Fourier domain.
                \\
                Visibility dynamic cross-spectrum & $V_{i,j}\left(\nu,t\right) = W_{i}\left(\nu,t\right)W^*_{j}\left(\nu,t\right)$ & The visibility of the pulsar between stations i and j at frequency $\nu$ and time $t$.
                \\
                Visibility conjugate spectrum  &
                $\widetilde{V}_{i,j}\left(\tau,f_D\right) = \left(\widetilde{W}_{i}* \widetilde{W}^*_{j}\right)\left(\tau,f_D\right)$ &  The Fourier transform of the visibility between stations i and j.
                \\
                Visibility secondary cross-spectrum &
                $S_{Vi,j}\left(\tau,f_D\right) = \widetilde{V}_{i,j}\left(\tau,f_D\right) \widetilde{V}_{i,j}\left(-\tau,-f_D\right)$ & Sensitive to the sum of the angular separations of images from the pulsar.
                \\
                \hline
            \end{tabular}
            \label{tab:definitions}
    \end{table}
\end{landscape}
\section{Data}
\label{sec:Data}
For a test case we use data taken for PSR B0834+06 by \cite{Brisken2009} in 2005. Data was taken simultaneously using Arecibo (AR), the Green Bank Telescope (GB), Jodrell Bank (JB), and tied-array Westerbork (WB). Visibilities and dynamic spectra with 244 Hz frequency resolution and $1.25$ s integrations were created for a $32$ MHz band centered around $316.5$ MHz and 110 minutes of simultaneous observation with the DiFX software correlator. Unfortunately, calibration issues with the WB data prevented its use, and only AR,GB, and JB were included in this analysis.

\subsection{Normalization}
In order to use the composite $\theta\mhyphen\theta$ method the data from different stations has to be renormalized. The first step is to apply a singular value decomposition (SVD) on each dynamic spectrum in order to divide out pulse to pulse variation and the band pass and set the mean to unity. Since the eigenvalue decomposition of the $\theta\mhyphen\theta$ matrix assumes uniform noise the dynamic spectra are then rescaled to have the same noise as the highest signal to noise spectrum (the Arecibo dynamic spectrum in this case). To do this, noise in measured in the secondary spectra using points far from the main arc. Each dynamic spectrum is then divided by the ratio of its noise to the reference spectrum. However, this also rescales the magnification vector for each dynamic spectrum. Since these vectors must be consistent between the dynamic spectra and visbilities, the visibilities must also be rescaled. As for the dynamic spectra, an SVD is used to remove pulse to pulse variation and the band pass. The SVD is also  used to correct lingering phase calibration issues in the visibility by removing large scale phase structures such that the average phase over many scintles is zero. This is done by derotating the visiblity at each point by the phase of the single mode SVD model (the $m=1$ term in Eq.~\ref{eq:svd_full}), similar to what is done in \cite{Simard2019}  Finally, each visibility is rescaled such the mean of its amplitude squared is equal to the mean of the product of the corresponding dynamic spectra as
\begin{equation}
    V_{\rm{normalized},i,j} = V_{i,j}\sqrt{<I_i I_j>/<|V_{i,j}|^2>}
\end{equation}

\section{VLBI phase retrieval of B0834+06}
\begin{figure}
    \centering
    \includegraphics{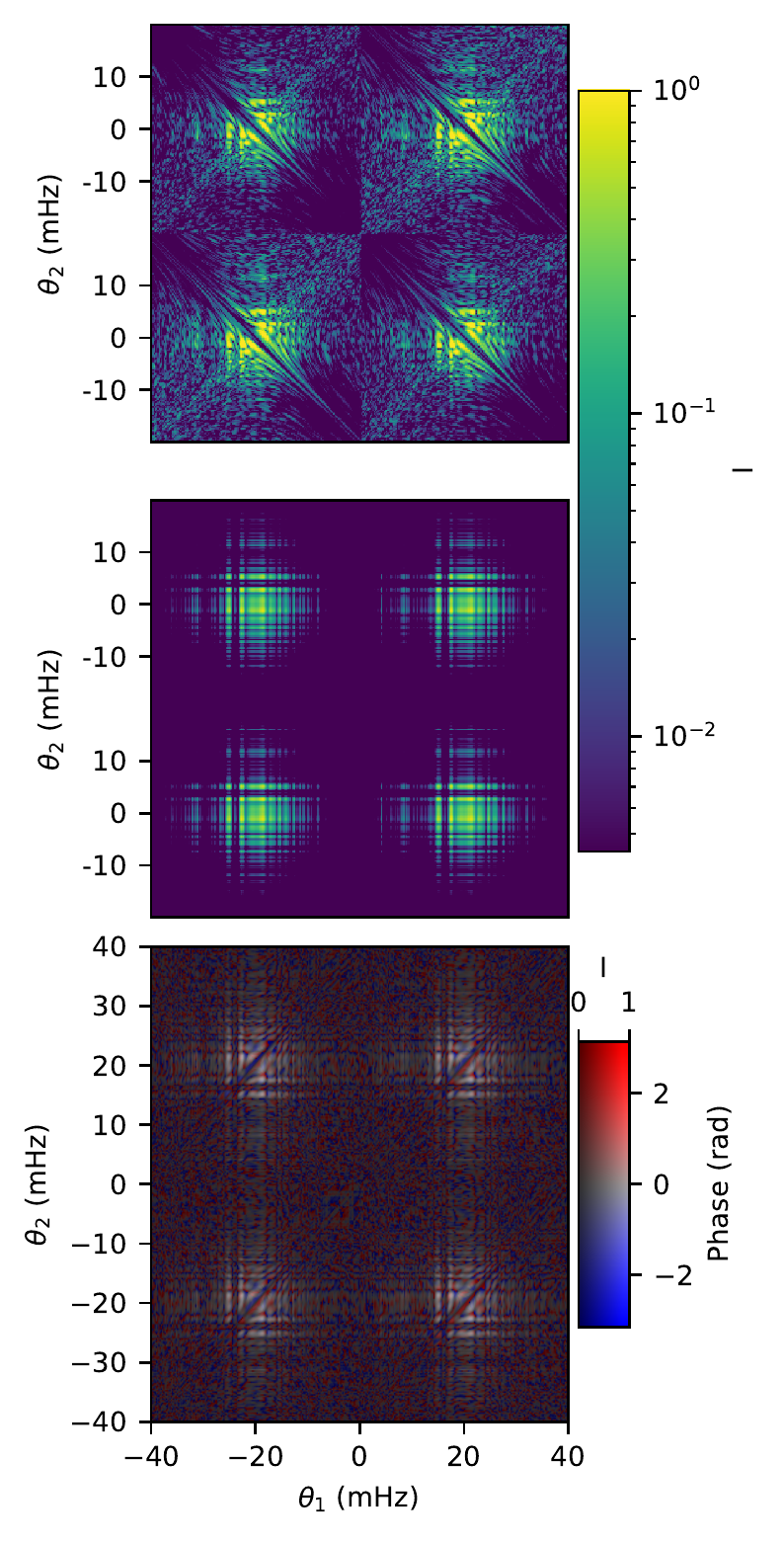}
    \caption{Magnitude squared of Data (top) and Model (middle) $\theta\mhyphen\theta$ spectra, as well as the phase difference between the two (bottom), for two station VLBI with Arecibo and Green Bank telescopes. For the phase difference, phases are shown by colour and Intensity relative to the max in saturation. Clockwise from top left the blocks of the matrix are made from the Arecibo dynamic spectrum, the visibility between the two stations, the Green Bank dynamic spectrum, and the complex conjugate of the visibility.}
    \label{fig:vlbi_2d}
\end{figure}

\begin{figure*}
    \centering
    \includegraphics{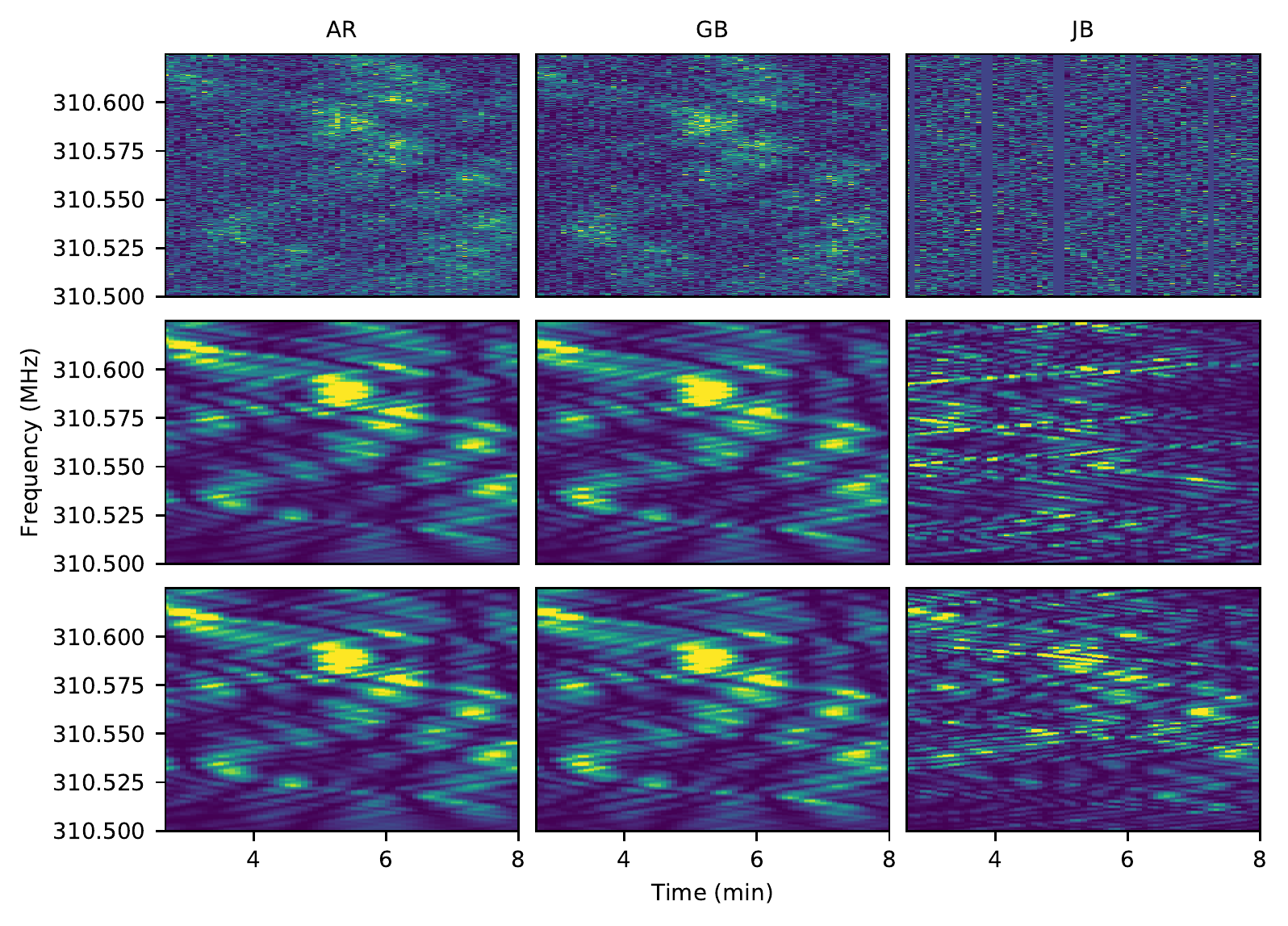}
    \caption{Comparison of the measured dynamic spectra (top) to the single dish (middle), and VLBI (bottom) models for three dishes. The high signal to noise ratio of the AR-JB and GB-JB visibilities yields a marked improvement in the JB recovery.}
    \label{fig:vlbi_dspec_comp}
\end{figure*}

The method for phase retrieval in the VLBI case remains much the same as for the single dish case. First, the data is divided into a series of chunks consisting of 10.5 minutes of data over 0.125MHz chosen such that each chunk overlaps halfway in time and or frequency with the chunks around it.  Next, the measured curvature of $5.401 \pm .003 \text{s}^3$ at $320 \text{MHz}$ from \cite{Baker2021} is scaled to the mean frequency of the chunk and used to generate $\theta\mhyphen\theta$ spectra for each dynamic spectrum and visibility. The $\theta\mhyphen\theta$ spectra from each individual dish and baseline are then combined into a single block matrix as described in Sec.~\ref{sec:theory}. An example of this using the two highest signal to noise dishes (Arecibo and Green Bank) is shown in Fig.~\ref{fig:vlbi_2d}. The dominant eigenvector of this matrix gives an estimate of the concatenated response vector defined in Eq.~\ref{eq:concat_u}. The outer product of $\bm{\mu}$ with itself is shown in the middle panel of Fig.~\ref{fig:vlbi_2d} for comparison purposes. The concatenated response is then broken up into the response vectors of the individual dishes. These are then mapped back into time-frequency space and the results from all chunks are combined using the mosaic approach described in \cite{Baker2021}. Briefly, the wavefield $W_n$ withing each chunk is added successively to the full wavefield $W$ such that within the overlapping region
\begin{equation}
    \arg\left(\left<W W_n^*\right>\right) =0
\end{equation}
However, to make sure that phases remain consistent between the dishes, we now used the weighted average of the overlapping regions at all dishes to determine the phase rotation when adding a new section. A portion of the dynamic spectra produced from the recovered wavefields using this method on AR,GB, and JB are show in Fig.~\ref{fig:vlbi_dspec_comp}. For comparison purposes we also include the results from single dish recovery at each station. Many features of the JB wavefield that were lost due the lower signal to noise of the dynamic spectrum are recovered through the information in the visibilities.
\label{sec:phase_ret}

\subsection{Recovery of the Millisecond Feature}
Since the millisecond feature is clearly not part of the same one dimensional structure as the main arc, it cannot be modeled directly with $\theta\mhyphen\theta$. However, the total power is dominated by the main screen which is also responsible for the inverted arclets at the millisecond feature. The observed feature in secondary spectrum is the convolution of the main arc with a collection of images near $1~\rm{ms}$. If we can recover the main arc then the millisecond feature can be recovered by any method that undoes the convolution. To this end, we perform the mosaic method described above using $\theta\mhyphen\theta$ to recover the wavefield only out to approximately $512~\mu\rm{s}$. We then employ an iterative approach similar to the Gerchberg-Saxton algorithm from \cite{Gerchberg72}: alternating between forcing the amplitudes of the wavefield to be the square-root of the dynamic spectrum, since this is a nearly direct measurement of the amplitude already, and setting all modes where $\tau<0$ in the conjugate wavefield to zero to enforce a causality constraint. The only significant difference between this and the original algorithm is that no amplitude forcing is applied to the the positive $\tau$ portion of the conjugate wavefield.

\section{Lens Parameters}
    \label{sec:lens_params}
\begin{figure}
        \centering
        \includegraphics{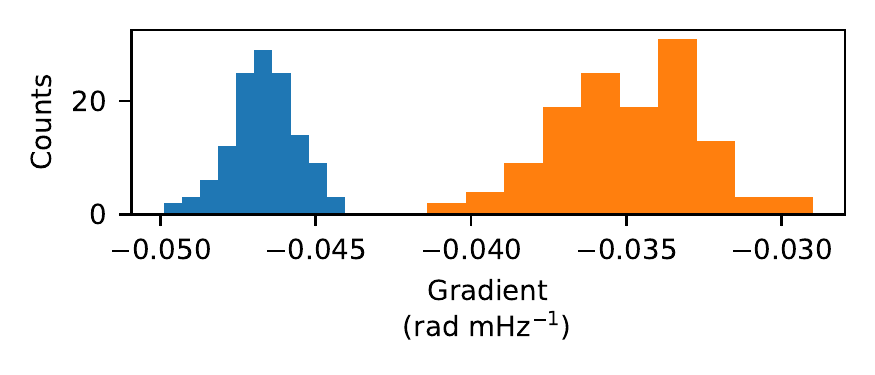}
        \caption{Distribution of best fit phase gradients using 0.25 MHz subbands of the ARGB (blue) and ARJB (orange) baselines. }
        \label{fig:main_grads}
    \end{figure}

    Since the main arc and millisecond feature appear distinct in the conjugate wavefield, we address the question of whether they come from distinct screens on the sky. We begin by modelling the main arc as a single line of images located at some effective distance $d_{\rm{eff},1}$ and rotated by an angle $\Omega_1$ east of the declination axis. For an image with a given time delay $\tau$, its angular offset is given by
    \begin{equation}
        |\theta| = \sqrt{\frac{2 c \tau}{d_{\rm eff,1}}}
    \end{equation}
    Since the sign of $f_D$ for an image determines which side of the line of sight it lies along the line of images, we use this to specify the side of the lens with
      \begin{equation}
        \theta(f_D,\tau) = \pm\sign{(f_D)}\sqrt{\frac{2 c \tau}{d_{\rm eff,1}}}
    \end{equation}
    the choice of $\pm$ will define which direction along the screen is positive and will rotate the value of $\Omega_1$ by 180 deg. We have chosen the $-$ convention as this gives a position angle between 0 and 180 deg.
    For a given chunk of our observation, centered at wavelength $\lambda_i$, we produce the cross conjugate wave-spectrum between two stations with projected baseline $\bm{b}$. The phase, $\phi$ for an image with Doppler frequency $f_D$ along the the main arc is given, assuming the angular offset of the image is small, by
    \begin{equation}
        \phi(f_D) = -2\pi \left[f_D\bm{b}\cdotp \bm{s}_1\sqrt{\frac{2 c
              \eta_i}{d_{\rm eff,1}}} \right] {\lambda_i^{-1}}
    \end{equation}
    where $\bm{s}_1$ is the unit vector along the screen and $\eta_i$ is the arc curvature at $\lambda_i$.
    Since $\eta$ is proportional to the square of the wavelength, this gives us a constant phase gradient over $f_D$ when our
    angular offsets are small (as is the case here) of $ -2\pi
    \bm{b}\cdotp \bm{s}_1{\sqrt{\frac{2 c \eta}{d_{\rm eff,1}}}}\lambda^{-1}$. If we rotate the cross conjugate wave-spectrum for this baseline by $-\phi(f_D)$, the signal should become entirely real and so the imaginary part will be entirely noise. Hence, our best fit gradient is the one that minimizes the sum of the squares of the imaginary part when used to derotate the data. The best fit phase gradients using 128 subbands of 0.25 MHz are shown in Fig.~\ref{fig:main_grads}. The average gradients for the two baseline are $-0.0479 \pm 0.0001~\rm{rad}~\rm{mHz}^{-1}$ for ARGB and $-0.0350 \pm 0.0002~\rm{rad}~\rm{mHz}^{-1}$ for ARJB. Together, these imply an effective distance of $d_{\rm{eff,1}} = 1.186\pm0.005~\rm{kpc}$ and position angle of $\Omega_1 = 155.0 \pm 0.1 \degr$

\begin{figure}
        \centering
        \includegraphics{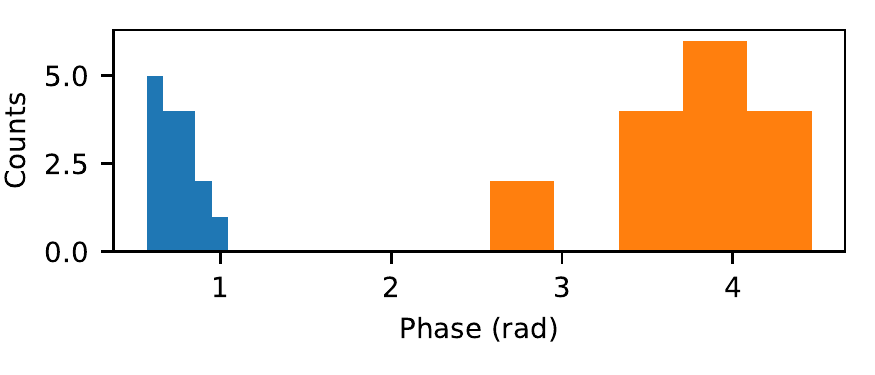}
        \caption{Distribution of fitted values of the phase at the apex of the millisecond feature using a fixed $\tau_0$ for ARGB (blue) and ARJB (orange) baselines scaled to $320~\rm{MHz}$. The phases of the ARJB baseline have been corrected by $\pi$ whenever the degenerate solution was found.}
        \label{fig:msf_phi0_fits}
    \end{figure}

    As previously observed in \cite{Brisken2009}, the images of the millisecond feature are offset from the main lens the line of images does not intersect our line of sight to the pulsar. As such we include the offset of the line of images as a third parameter in our model. Unlike \cite{Liu2016}, our model assumes that each image is the result of lensing by single screen with no doubly lensed images. We have chosen this as the mathematically simplest model of the scattering, though not necessarily the physically simplest. It should also serve to disprove the hypothesis of both screens being at the same distance as double scattering should be virtually impossible in that case. If the data is incompatible with the two lenses being at the same distance in this simple picture it should be generally ruled out. A more complete discussion of double lensing and possible structures that might be responsible is given in \cite{Zhu2023} and \cite{Jow2023} For a line of images at effective distance $d_{\rm{eff,2}}$, whose closest point to the line of sight is offset from the line of sight by $\bm{\theta}_0$ we model the phase difference along a baseline $\bm{b}$ as

    \begin{equation}
        \phi(\tau) = \pm 2\pi\left(\bm{b} \cdotp \bm{\theta}_0 + \sqrt{\frac{2 c \tau}{d_{\rm{eff,2}}}-\left|\bm{\theta}_0\right|^2}(\bm{b} \cdotp \bm{s}_2)\right)\lambda^{-1}
        \label{eq:msf_grad}
    \end{equation}

    where $\bm{s}_2$ is the unit vector along the lens which is perpendicular to $\bm{\theta}_0$. The sign of the phase will depend on which direction along the lens we define to be positive as well as which side of the closest point the image in question lies on. We define three parameters to describe the phase at closest approach, $\phi_0$, phase evolution, $B$, and minimal time delay, $\tau_0$ for the millisecond feature as
    \begin{eqnarray}
    \phi_0  &=& \frac{2\pi\bm{\theta}_0\cdotp\bm{b}}{\lambda}\\
    B &=& \frac{2\pi\bm{s}_2\cdotp\bm{b}}{\lambda}\sqrt{\frac{2c}{d_{\rm{eff}},2}} \\
    \tau_0 &=& \frac{d_{\rm{eff},2}}{2c} \theta_0^2
    \end{eqnarray}
    which simplifies Eq. \ref{eq:msf_grad} to
        \begin{equation}
        \phi(\tau) = \pm \left(\phi_0 + B \sqrt{\tau - \tau_0} \right)
    \end{equation}
    The most important of these parameters for our work are $\phi_0$, which gives the relative astrometric position of the closest approach, and $\tau_0$ which can be combined with the astrometric position to determine effective distance. Using a fixed value of $\tau_0 = 954.6~\mu\rm{s}$ from the observations of \cite{Zhu2023} who can track its motion over several observations and interpolate to this observation.  we fit for the remaining parameters along each baseline by minimizing the sum of the squares of the imaginary parts after derotation by our model. There are technically two families of solutions to this minimization offset by $\pi~\rm{rad}$ corresponding to either a negative or positive real part. In our case the positive solution is correct as that would correct the phases at one dish to be the same as the other. In cases where the negative solution is found we add $\pi$ to $\phi_0$. To choose a solution within a family, we add or subtract a multiple of $2\pi$ to $\phi_0$ in each band until they are within $\pi$ of the first band. Since the millisecond feature is fainter than the main arc, wider frequency bands are used for the fitting in order to increase the signal to noise ratio. The results of these fits for $\phi_0$ for the sixteen $2~\rm{MHz}$ subbands across the observation are show in Fig.~\ref{fig:msf_phi0_fits} after rescaling to $320~\rm{MHz}$.Averaging the scaled values of $\phi_0$ gives final values of $\phi_0 = 0.73 \pm 0.03~\rm{rad}$ for the ARGB baseline and $\phi_0 = 3.8 \pm 0.1~\rm{rad}$ for the ARJB baseline at $320~\rm{MHz}$. Since the derotation is equally effective with the addition of any multiple of $2\pi$ to either baseline we have chosen our $\phi_0$ such that the variance in the astrometric position of the apex of the feature across all bands is minimized. Mapping onto the sky give $d_{\rm{eff},2} = 1.4 \pm 0.1~\text{kpc}$, $\theta_0 = 23.7 \pm .8~\rm{mas}$, and a position angle towards the closest point of $\Omega_2 = 37 \pm 1\degr$. All errors assume the measured curvature from \cite{Baker2021} is exact since the fractional error of less than 0.01 percent is much smaller than the errors in the phase of the conjugate wavefields.

\begin{figure}
        \centering
        \includegraphics[width=\columnwidth]{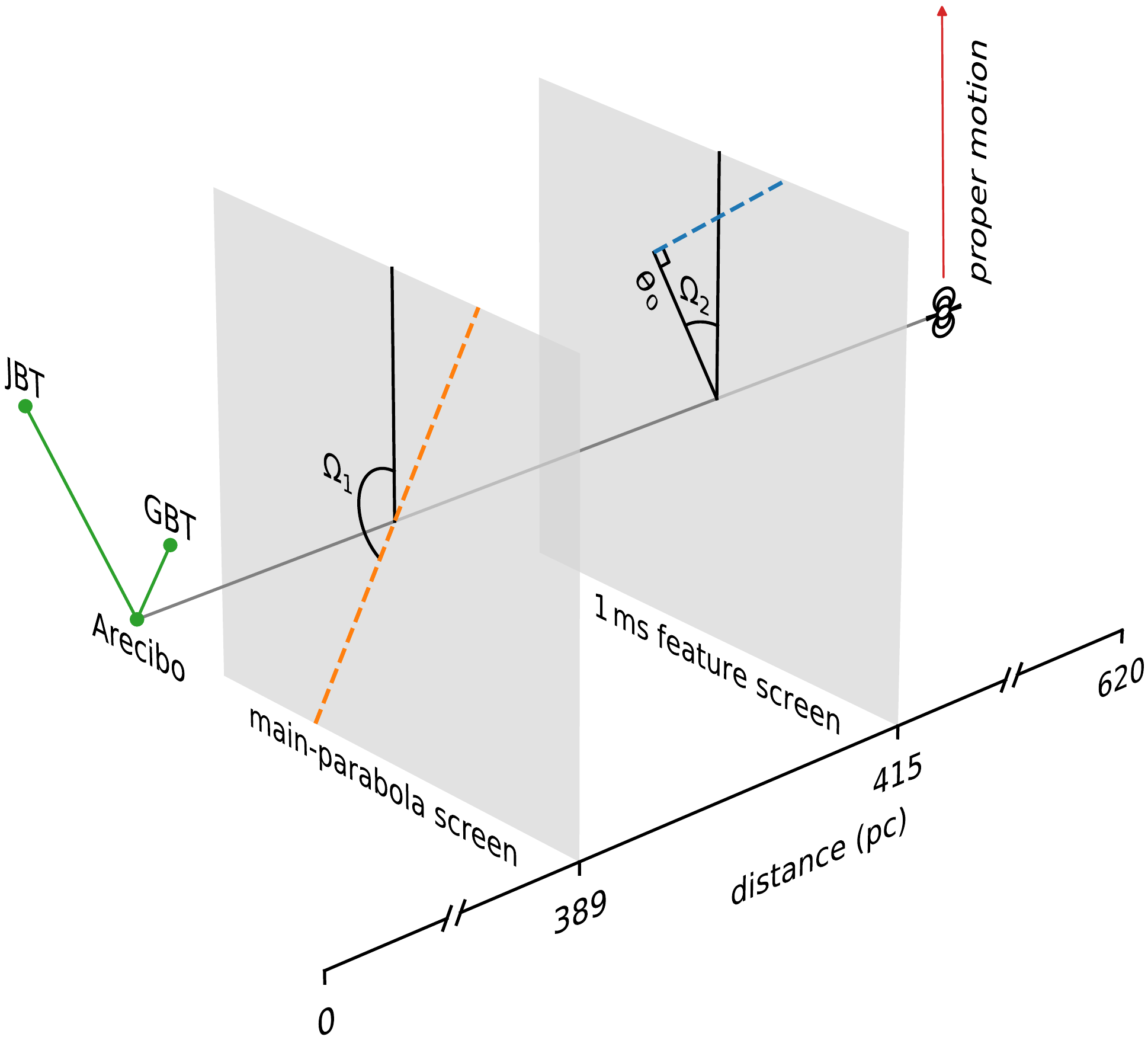}
        \caption{A simple schematic of the two screens. The dotted orange line indicates the orientation of the main screen, while the blue line shows the lens responsible for the millisecond feature. Since features are only seen on one side of the closest point we only show that side of the millisecond lens. The angles $\Omega_1$ and $\Omega_2$ as well as the closest offset $\theta_0$ used to define the lenses are also shown. Green lines show the two baselines used in our analysis. Distances are shown assuming the fiducial pulsar distance of 620 pc}
        \label{fig:simple_schematic}
    \end{figure}

\begin{figure*}
        \centering
        \includegraphics{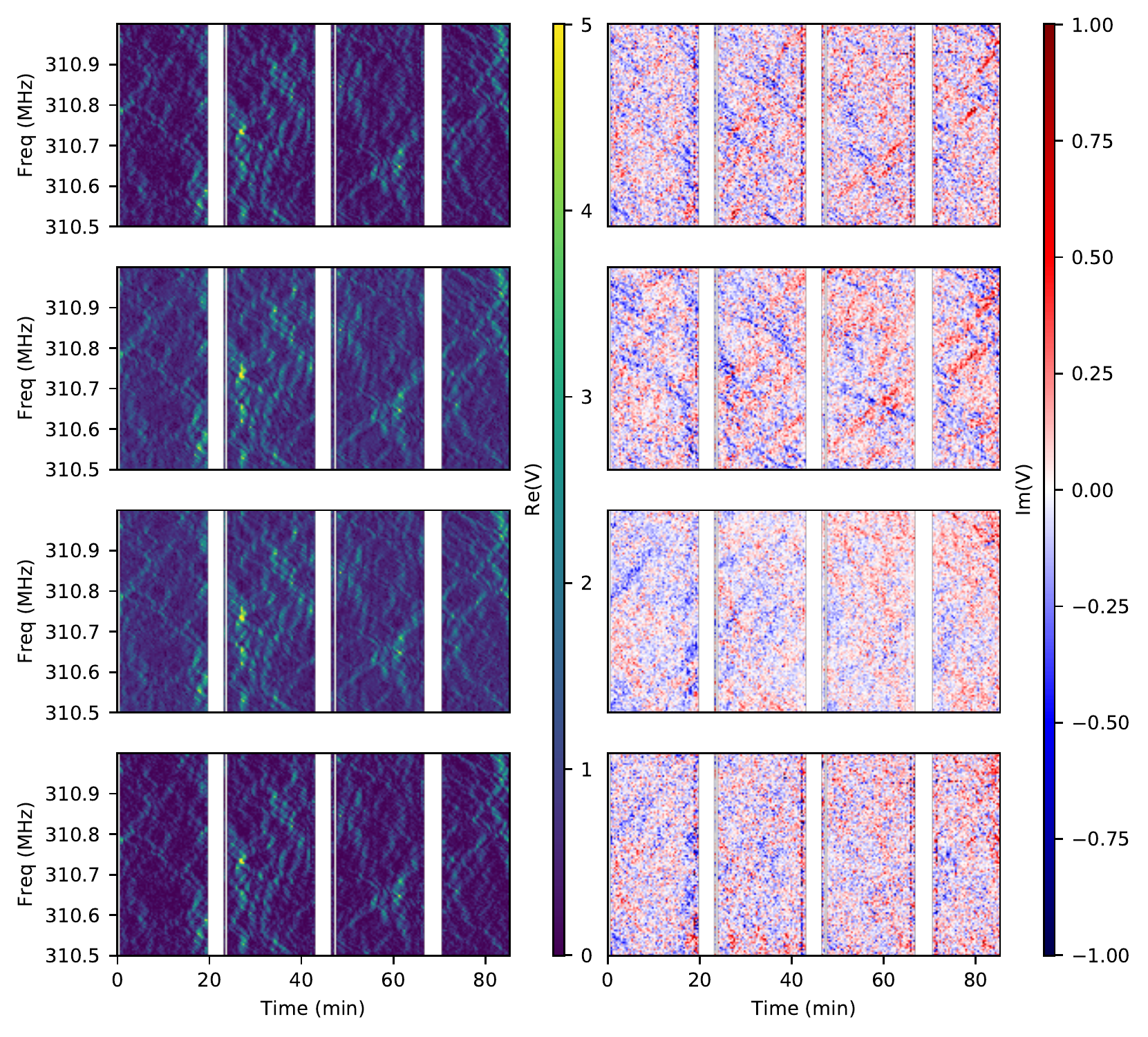}
        \caption{Sample region of the real (left) and imaginary (right) parts of ARGB visibility during different stages of modelling. The raw data (top) and $\theta\mhyphen\theta$ model (second from top) show a clear crisscross structure due to power on different sides of the parabola. After correcting for the phase difference between the dishes from our five parameter model (second from bottom) most of this structure is removed. Finally, the same phase difference is applied to the raw data (bottom) and we see that our model is able to account for almost all of the imaginary structure.}
        \label{fig:vis_mod}
    \end{figure*}

    A simple schematic of the two screens projected onto the sky is shown in Fig.~\ref{fig:simple_schematic}. In order to test our model examine the ARGB baseline through different stages of modelling in Fig.~\ref{fig:vis_mod}. A key feature of the data visibility is the crisscross pattern of features in the imaginary part. The negative features, running down and to the right, and the positive features, running down and to the left, come from features on different sides of the parabola in the conjugate spectrum. The signs of the crisscross are due to the time delay between the dishes. If we can correctly model this delay, we should be able to remove the imaginary features and leave only the real part. At that point, we have extracted all the new information from VLBI and are left with only the dynamic spectrum. Initially, these features are also quite clear in the $\theta\mhyphen\theta$ model of the visibility, which is a good indication of its success at capturing the data. Using the five parameter fit described above, we can derotate the Green Bank conjugate wavefield to remove the effects of the time delays to different images relative to Arecibo. Transforming back to a wavefield and generating a new model visibility with Arecibo removes much of the imaginary structure. This also allows us to see how the model affects the phase at each point in the visibility. Applying the same change of phase to the raw data also removes most of the imaginary part.

\begin{table}
     \caption{Measured and inferred parameters for the main and millisecond feature lenses.}
    \centering
    \begin{tabular}{||c|l||}
    \hline
    Parameter & Value \\
    \hline
    \multicolumn{2}{c}{Measured (Scaled to 320 MHz)}\\
    \hline
    ARGB Main Arc Gradient (rad $\rm{mHz}^{-1}$) & $-0.0467\pm0.0001$\\
    ARJB Main Arc Gradient (rad $\rm{mHz}^{-1}$) & $-.0350\pm0.0002$\\
    $\phi_{0,ARGB}$ (rad) & $0.73\pm0.03$\\
    $\phi_{0,ARJB}$ (rad) & $3.8\pm0.1$\\
    \hline
    \multicolumn{2}{c}{Inferred}\\
    \hline
    $d_{\rm{eff},1}$ (kpc) & $1.186\pm.005$\\
    $d_{\rm{eff},2}$ (kpc) & $1.4\pm.1$\\
    $\Omega_1$ (deg) & $155.0\pm0.1$ \\
    $\Omega_2$ (deg) & $37\pm1$ \\
    $\theta_0$ (mas) & $23.7\pm.8$ \\
    \end{tabular}
\end{table}

\subsection{Examining Residuals}
\begin{figure}
    \centering
    \includegraphics{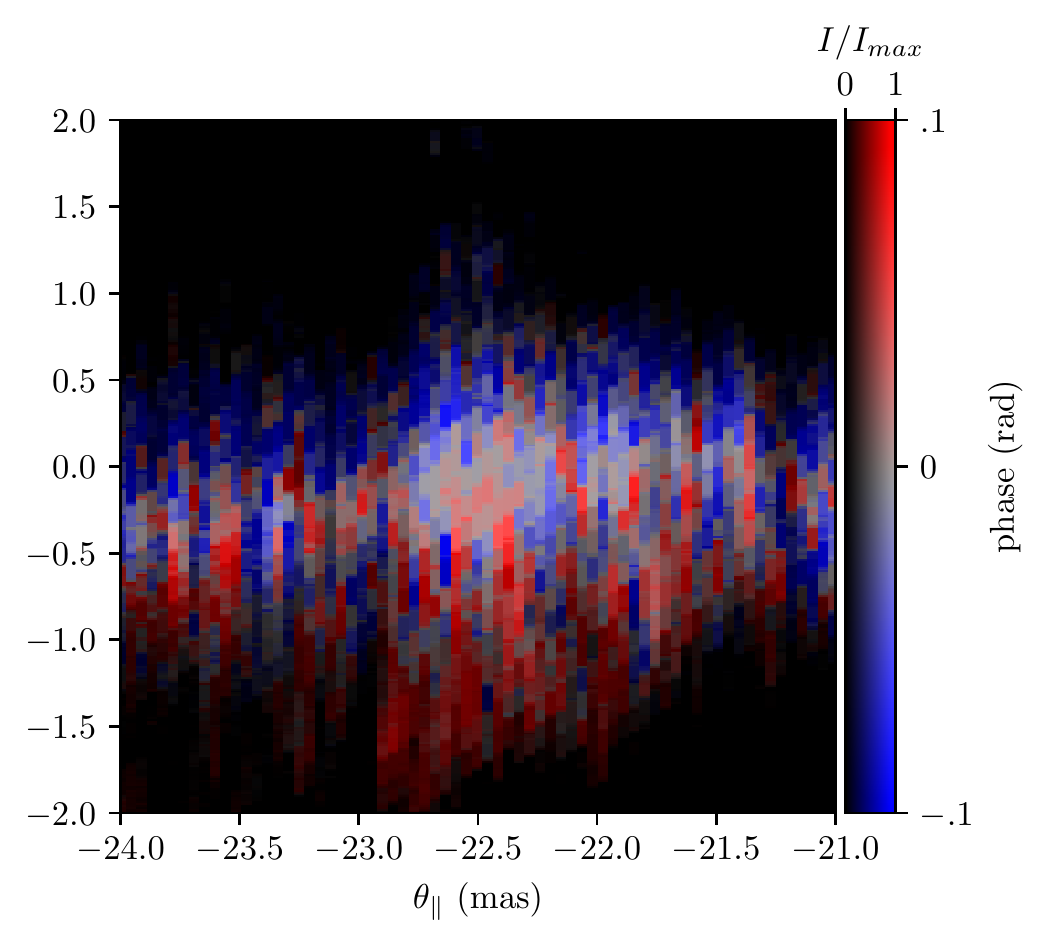}
    \caption{Phase of the $\theta\mhyphen\theta$ spectrum near the 300 $\mu$s feature after removing our best fit model with lightness used to indicate relative intensity. We have labeled the axes with $\theta_\perp$ and $\theta_\parallel$ to help relate them to coordinates along and perpendicular to the main axis of the screen.}
    \label{fig:300us}
\end{figure}

After removing the phase gradient from our best fit physical parameters, we can look for areas in the cross conjugate wave-spectrum or its $\theta\mhyphen\theta$ transform that deviate from the model. Since signals are more spread out in the conjugate spectrum than the conjugate wavefield, it is easier to locate these small deviations. 
As discussed \cite{Baker2021}, a small feature can be seen to branch from the main arc at approximately $300~\mu\rm{s}$ on the negative Doppler side of the arc. Examining this feature in $\theta\mhyphen\theta$ after the removal of our model phase gradient, we see, in Fig. \ref{fig:300us}, that there exists some residual phase offset on the feature on the ARGB baseline. In this space the feature deviates from the main arc near $-21.5~\rm{mHz}$ along the main axis ($\theta_\parallel$) moving towards negative $\theta_\perp$. Since this feature is at the same time delay as the main arc, this phase offset may indicate a slightly different distance. Unfortunately, the ARJB baseline is too noisy in this region to allow us to properly map the feature to determine its distance.

\begin{figure}
    \centering
    \includegraphics{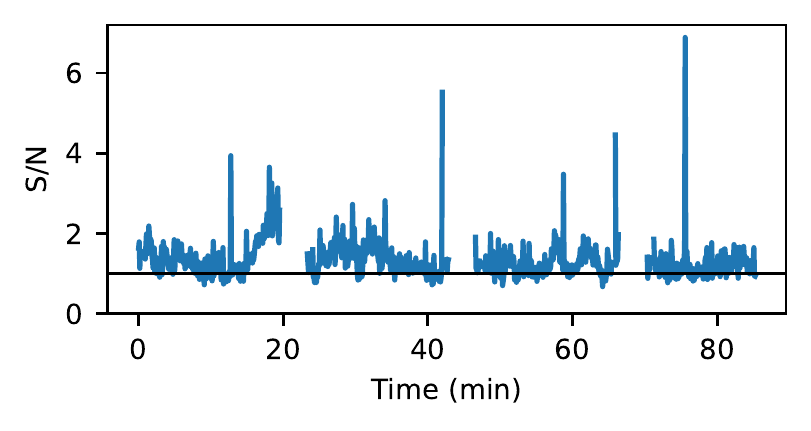}
    \caption{Residual power to noise power ratio as a function of time. We expect to see the ratio increase near the ends of the observation and gaps where the $\theta\mhyphen\theta$ model is mose likely to break down.}
    \label{fig:respow}
\end{figure}

We expect that any systematic errors in our results arise from any regions in the wavefield not accurately modeled by $\theta\mhyphen\theta$ and the correction for double lensing in the millisecond feature. $\theta\mhyphen\theta$ is most likely to run into problems at the edges of the data, where there are fewer chunks overlapping, and around gaps where chunks have less data to work with. We can see some signs of this in Fig. ~\ref{fig:vis_mod} where features in the residual imaginary component tend to cluster near the edges of the data and gaps. Fig.~\ref{fig:respow} shows residual power to noise power ratio for each time bin in the imaginary part of the derotated visibility shown in Fig~\ref{fig:vis_mod}. The residuals are generally consistent with the noise except for a few isolated time bins and a noticeable deviation near the first gap. The S/N ratio of the entire field is approximately $1.6$. 

\section{Astrometric Imaging}
\label{sec:astromet}
\begin{figure*}
    \centering
    \includegraphics{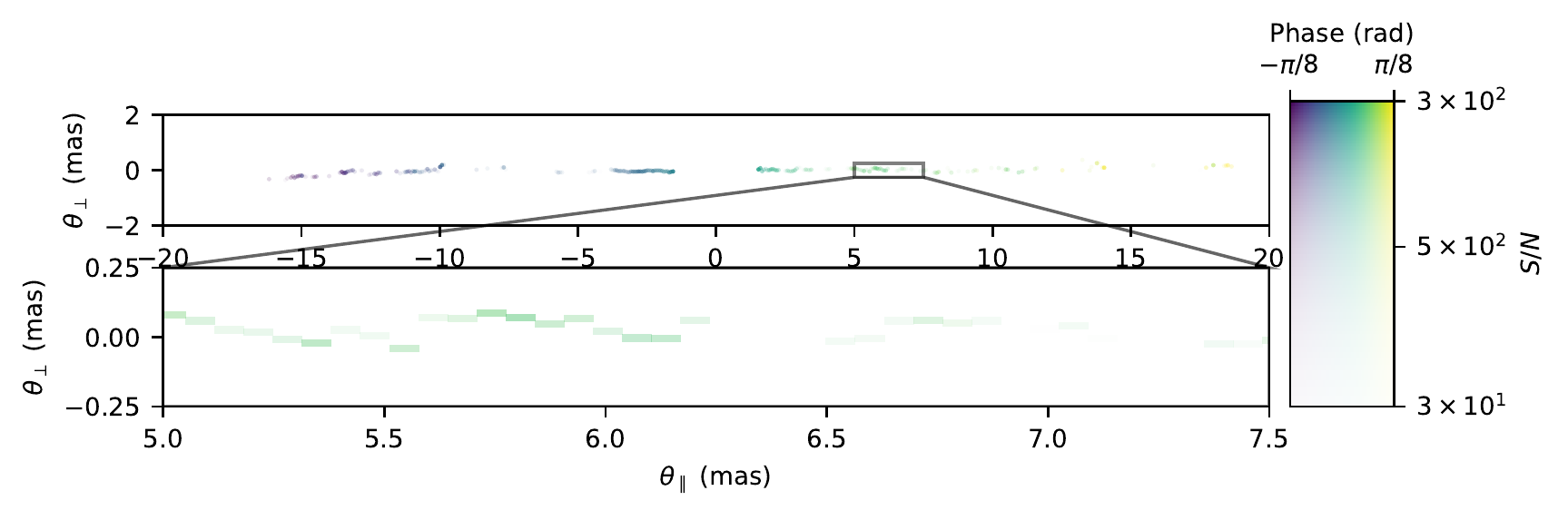}
    \caption{Offset of peak brightness perpendicular to the arc ($\theta_\perp$) as a function of distance along the major axis of the lens ($\theta_\parallel$). The lens is extremely anisotropic with an axial ratio greater than 60. The colour scale shows the residual phase of these points after removing our best fit gradient.}
    \label{fig:parper}
\end{figure*}

In addition to the general lens parameters described above, VLBI $\theta\mhyphen\theta$ methods also allow for high precision relative astrometry of the lens. In particular, we are able to search for any deviations from our one dimensional model. By applying the $\theta\mhyphen\theta$ transformation to the recovered cross conjugate wave-spectrum for ARGB in 2MHz bands, after rescaling $f_D$ to correct curvatures to 320 MHZ, and averaging over all subbands, we create a map of the image positions in coordinates parallel and perpendicular to the major axis of the screen.  Since the $\theta\mhyphen\theta$ transform has coordinates given in $f_D$, the $d_{\rm{eff,1}}$ from our global fits is used to convert to mas on the sky. For each value of $\theta$ along the main arc, we measure the offset and phase of the brightest point in $\theta_\perp$. This approach is similar to the back-mapped astrometry described in \cite{Brisken2009}. The results are shown in Fig. ~\ref{fig:parper}. As we get to larger values of $\theta_\perp$, our signal to noise ratio per point drops and so we see a larger spread in the peak position as well as variations from our phase gradient.
For $\theta_\parallel$ between -15 and 15 mas, 95 percent of measured peaks fall within 0.24 mas of the major axis suggesting an axial ratio of greater than ~60.

\begin{figure}
    \centering
    \includegraphics{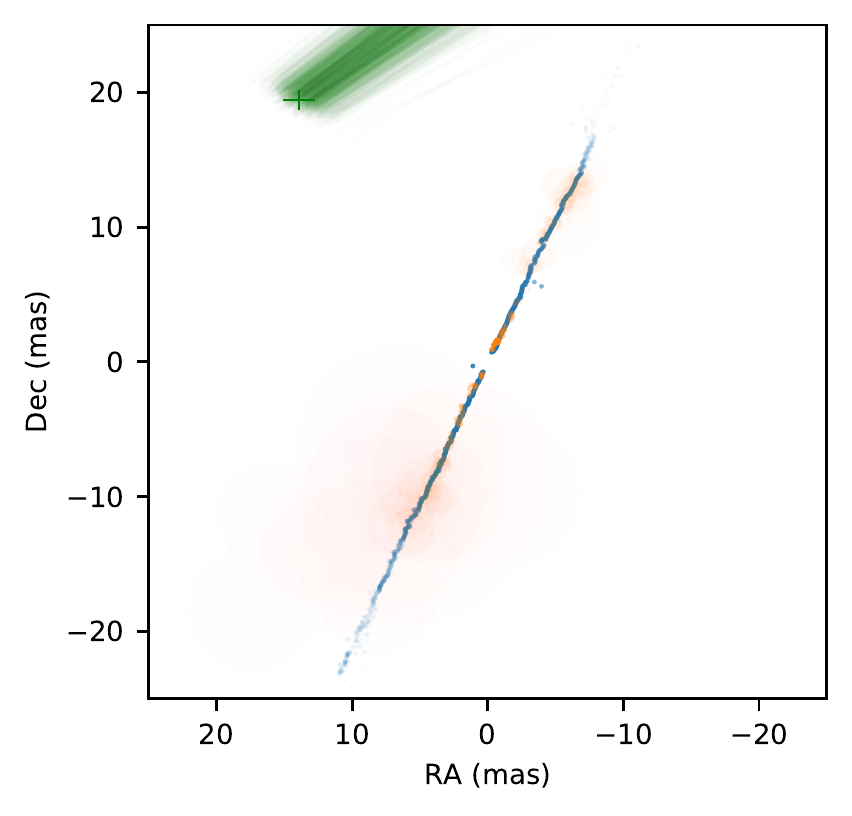}
    \caption{VLBI positions of 165 images along the main arc (orange) and the offsets measured from the frequency averaged $\theta\mhyphen\theta$ along the screen direction measured in Sec.~\ref{sec:lens_params} (blue). With the relative brightness of points indicated with opacity and the size of the orange points indicating the uncertainty on the VLBI image position The millisecond feature is shown in green with error bars for the point of closest approach and probability of the line of images passing through each point shown with opacity}
    \label{fig:skymap}
\end{figure}

To examine individual images directly from VLBI, 165 isolated peaks along or near the main arc were identified by eye from the frequency averaged $\theta\mhyphen\theta$ diagram. The average phase in a small region about each point was measured for the ARGB and ARJB cross conjugate wave-spectra in each of the 2MHz subbands in order to measure its relative astrometric position. The average position and standard error for each of these points is shown in Fig.\ref{fig:skymap}.
As noted in \cite{Brisken2009}, the back-mapped astrometric positions scatter less than those from VLBI phase measurements. Errors on the VLBI positions are on the order of 0.6 mas which makes them too large to detect the variations on the order of 0.2 mas seen from the back-mapping. 

A crucial difference in our astrometric results compared to \cite{Brisken2009} is that the location of the millisecond feature is in the upper left quadrant as opposed to the lower left. In fact, all images have been mirrored in both right ascension and declination. This picture with the millisecond feature in the north is consistent with the monitoring results of \cite{Zhu2023}  who see the time delay of the feature decreasing during this epoch suggesting that the pulsar (moving almost due north) is approaching the feature.  One possible explanation for this flip is that the choice of Fourier transform convention used to generate the spectra from baseband data can result in producing the complex conjugate of the visibilities, which in turn results in a mirroring of the VLBI images. $\theta\mhyphen\theta$ allows us the make an additional test of the visibilities in cases such as this where a comparison of the response functions from single dish recovery can be compared to the results from visibilities alone. As seen in Fig.~\ref{fig:visibility_veccomps}, when the consistent choice of visibility conjugation is chosen each dish sees only a constant phase offset between the two methods. However, when the complex conjugate of the visibility is used to make the $\theta\mhyphen\theta$ matrix axes corresponding to the two dishes are reversed and the comparison will now produce a phase gradient determined by the time delay between the two dishes. 

We attribute the improvement in our measured phases, and hence astrometric results, to the collapse of arclets back to a single point for each image. Not only does this make it easier to locate individual points, but we expect that it increases the signal strength at that point due to the deconvolution. In effect we have taken the coherent sum over the entire arclet instead of taking only the information from the apex. This can be seen by taking the ratio of the average power along the main arc to the average power far away in a noise dominate region. For the Arecibo secondary spectrum this ratio is approximately $30$ whereas for the Arecibo secondary wavefield it is approximately $780$.

\section{Ramifications}
Using new $\theta\mhyphen\theta$ techniques, we are able to measure the effective distance to the two screens seen for PSR B0834+06 to be $1.186 \pm 0.005~\rm{kpc}$ and $1.4 \pm 0.1~\rm{kpc}$ for the main arc and millisecond feature respectively. For the main arc, our results are consistent with the $1.171\pm0.023~\rm{kpc}$ reported by \cite{Brisken2009}.  They also measure a slightly larger distance to the millisecond feature, but as it is within one sigma of the main arc effective distance they conclude the lenses are at the same distance.

The success at retrieving the dynamic wavefields from all dishes in this well understood case also serves as a template for future work. Of particular interest is the potential to apply these methods to Earth-Space baselines such as RadioAstron. A natural starting ground for this would be archival data of PSR B0834+06 as described in \cite{Smirnova2020}. The observations taken in 2015 offer the most promising path as they include more ground stations which improves the ability of $\theta\mhyphen\theta$ methods to recover the wavefield of the low signal to noise RadioAstron data. The projected baseline for RadioAstron to Arecibo is approximately eighty times longer than Arecibo to Green Bank. The resulting increase in resolving power should allow us to improve our VLBI astrometry precision to below the 0.2mas level deviations indicated by the back-mapping.
\section*{Acknowledgements}
We acknowledge the support of the Natural Sciences and Engineering Research Council of Canada (NSERC), [funding reference number RGPIN-2019-067, 523638-201]. We receive support from Ontario Research Fund—research Excellence Program (ORF-RE), Simons Foundation, Canadian Institute for Advanced Research (CIFAR), and Alexander von Humboldt Foundation. The Arecibo Observatory  is  operated  by  SRI  International  under  a  cooperative agreement with the National Science Foundation (AST-1100968), and in alliance with Ana G. M\'endez-Universidad Metropolitana,  and  the  Universities  Space  Research  Association.  The  National  Radio  Astronomy  Observatory  is  a facility of the National Science Foundation operated under cooperative agreement by Associated Universities, Inc. Computations were performed on the Niagara supercomputer at the SciNet HPC Consortium. SciNet is funded by: the Canada Foundation for Innovation; the Government of Ontario; Ontario Research Fund - Research Excellence; and the University of Toronto. This research made use of Astropy,\footnote{http://www.astropy.org} a community-developed core Python package for Astronomy \cite{astropy:2013, astropy:2018}.

\section*{Data Availability}
The code used in generating $\theta\mhyphen\theta$ diagrams as well as fitting for curvatures can be found as part of the scintools python package originally developed by Daniel Reardon at \href{https://github.com/danielreardon/scintools}{github.com/danielreardon/scintools} (\cite{reardon2020precision}).



\bibliographystyle{mnras}
\bibliography{example.bib} 




\appendix


\bsp	
\label{lastpage}
\end{document}